\documentclass[12pt]{iopart}

\usepackage{iopams}

\usepackage{graphicx}
\usepackage{dcolumn}
\usepackage{float}
\usepackage{iopams}
\usepackage{multirow}

\begin{document}

\title{\boldmath Electron-Phonon Spectral Density of MgB$_2$ from Optical Data, through Maximum Entropy \unboldmath}

\author{Jungseek Hwang$^{1}$ and J. P. Carbotte$^{2,3}$}
\address{$^1$Department of Physics, Sungkyunkwan University, Suwon, Gyeonggi-do
440-746, Republic of Korea\\ $^2$Department of Physics and Astronomy, McMaster University, Hamilton, ON L8S 4M1, Canada \\$^3$The Canadian Institute for Advanced Research, Toronto, ON M5G 1Z8 Canada}

\ead{jungseek@skku.edu}

\date{\today}

\begin{abstract}
We use maximum entropy techniques to extract an electron-phonon density from optical data in the normal state at $T =$ 45 K in MgB$_2$. Limiting the analysis to a range of phonon energies below 110 meV which is sufficient to capture all phonon structures we find a spectral function which is in good agreement with that calculated for the quasi two-dimensional $\sigma$-band. Extending the analysis to higher energies up to 160 meV we find no evidence for any additional contributions to the fluctuation spectrum but find that the data can only be understood if the density of states is taken to decrease with increasing energy.
\end{abstract}

\pacs{74.25.Gz, 74.25.Jb, 74.25.-q}

\maketitle

\section{Introduction}

The discovery\cite{nagamastu:2001} of superconductivity in MgB$_2$ with a critical temperature $T_c =$ 39 K lead to several density functional LDA band calculations which showed strong coupling of the quasi two-dimensional $\sigma$-bands to optical B-B bond stretching phonons at 600 cm$^{-1}$\cite{kortus:2001,an:2001,liu:2001,Kong:2001,bohnen:2001,choi:2002,choi:2002a}. Evidence of strong coupling is that this band supports a large superconducting gap. A three dimensional $\pi$-band has a smaller gap so that MgB$_2$ is a multiband superconductor\cite{liu:2001,choi:2002,choi:2002a,golubov:2002,brinkman:2002,mazin:2002,dolgov:2003,nicol:2005}. The electron-phonon spectral density $\alpha^2F(\Omega)$ has been extracted from tunneling data in many conventional superconductor and used with considerable success to compute their properties\cite{carbotte:1990} which often differ significantly\cite{carbotte:1995,mitrovic:1980} from BCS predictions because of retardation effects. These are in addition to possible corrections coming from energy dependence in the electronic density of states\cite{mitrovic:1983:2} and anisotropies\cite{leavens:1971,odonovan:1995b}. It has also been possible to extract $\alpha^2F(\Omega)$ from optical data with excellent agreement with the tunneling results in the case of Pb\cite{farnworth:1976}.

In this paper we use optical data on the normal state of MgB$_2$ to get information on its electron-boson spectrum\cite{tu:2001,kuzmenko:2002}. To achieve this end we use maximum entropy techniques developed recently that have so far been used mainly to describe the electron-boson spectral density in the cuprates\cite{carbotte:2011} where the interaction is believed to be mediated through the exchange of spin fluctuations\cite{carbotte:2011,heumen:2009,heumen:2009b}. This work is based on a series of developments which started with the work of Allen\cite{allen:1971} who derived using ordinary perturbation theory, a simplified but very useful relationship between the optical scattering rate and the electron-boson spectral density. This work was later generalized to finite temperature by Shulga {\it et al.}\cite{shulga:1991} and by Mitrovic {\it et al.}\cite{mitrovic:1985} to include the possibility that the density of electronic states (DOS) has important energy dependence in the range of energy of interest. Further work by Sharapov and Carbotte\cite{sharapov:2005} included finite temperature as well as energy dependence in the density of states and derived simplified formulas which have turned out to be very helpful for analyzing optical data. These equations related the spectral density $\alpha^2F(\Omega)$ directly to the optical scattering rate through an integral equation which related $\alpha^2F(\Omega)$ linearly to the scattering through known kernel. The inversion problem involves a deconvolution which was addressed by Dordevic {\it at al.} using a singular value decomposition (SVD) technique\cite{dordevic:2005} and by Schachinger {\it et al.} using a maximum entropy technique (MET)\cite{schachinger:2006}. Here we apply this second method to invert the finite temperature optical data in MgB$_2$ at $T =$ 45 K above the superconducting transition temperature $T_c =$ 39.6 K. In section II we present the basic integral equation which we employ to relate the optical scattering rate to the spectral density $\alpha^2F(\Omega)$ which we wish to extract from the data. The necessary kernel which is known is specified and depends on the thermal factors and the underlying electronic density of states $N(\omega)$ which can have energy dependence. The maximum entropy technique used to solve for $\alpha^2F(\Omega)$ is specified in section III. Results are presented in section IV and conclusions given in section V.

\section{Formalism}

Sharapov and Carbotte\cite{sharapov:2005} have derived a formula for the optical scattering rate which is highly suitable for analysis of infrared data. The work is based on a Kubo formula without vertex corrections within a boson exchange model of correlation effects. Through a series of simplifications they obtain an integral equation which relates linearly through a known kernel $K(T,\omega,\Omega)$, the scattering rate $1/\tau^{op}(T,\omega)$ at temperature $T$ to the underlying spectral density $\alpha^2F(\Omega)$ which in our case is due to the electron-phonon interaction. The equation is
\begin{equation}\label{eq1}
\frac{1}{\tau^{op}(T,\omega)}=\int^{\infty}_{0} d\Omega \alpha^2F(\Omega)K(T,\omega,\Omega)
\end{equation}
The kernel $K(T,\omega,\Omega)$ is given by
\begin{eqnarray}\label{eq2}
K(T,\omega,\Omega) &=& \frac{\pi}{\omega}\int^{+\infty}_{-\infty}dz \tilde{N}(z-\Omega)[n_B(\Omega)+1-f(z-\Omega)]\nonumber \\ &\times &[f(z-\omega)-f(z+\omega)]
\end{eqnarray}
where $n_B(\Omega)$ and $f(\Omega)$ are the Bose-Einstein and Fermi-Dirac distributions at finite $T$, respectively. The density of states $\tilde{N}(\omega)$ is the particle-hole symmetrized version of the energy dependent band structure density of states $N(\omega)$ {\it i.e.} $\tilde{N}(\omega) = [N(\omega)+N(-\omega)]/2$. When $\tilde{N}(\omega)$ is assumed to be a constant independent of $\omega$ and is set equal to 1.0 we recover the formula of Shulga {\it et al.}\cite{shulga:1991} namely
\begin{eqnarray}\label{eq3}
\frac{1}{\tau^{op}(T,\omega)}&=&\frac{\pi}{\omega}\int^{\infty}_{0} d\Omega \alpha^2F(\Omega)\Big{[} 2\omega \coth \Big{(}\frac{\Omega}{2T} \Big{)} \nonumber \\
&-&\!(\omega\!+\!\Omega) \coth \!\! \Big{(}\!\frac{\omega\!+\!\Omega}{2T}\! \Big{)}\!+\!(\omega \!-\! \Omega) \coth \!\! \Big{(}\!\frac{\omega\!-\!\Omega}{2T}\! \Big{)}\Big{]}
\end{eqnarray}
which is related, but different, from the quasiparticle scattering rate $1/\tau^{qp}(T,\omega)$ given instead by the formula
\begin{eqnarray}\label{eq4}
\frac{1}{\tau^{qp}(T,\omega)}&=&\frac{\pi}{2}\int^{\infty}_{0} d\Omega \alpha^2F(\Omega)\Big{[} 2\coth \Big{(}\frac{\Omega}{2T} \Big{)} \nonumber \\
&-&\tanh \Big{(}\frac{\omega+\Omega}{2T} \Big{)}+\tanh \Big{(}\frac{\omega-\Omega}{2T} \Big{)}\Big{]}.
\end{eqnarray}
It is the approximate Eq. (\ref{eq3}) for the optical scattering rate that we use in all the inversion of optical data that we present here. The assumption of a constant density of states is acceptable for MgB$_2$ in the energy range of interest here. Further Eq. (\ref{eq3}) is known to be surprisingly a good approximation to the optical scattering rate obtained from full Eliashberg solutions of the electron-phonon system\cite{schachinger:2006}.

We will also be interested in including residual impurity scattering. As discussed in Sharapov and Carbotte, this corresponds to a spectral density $\alpha^2F_{imp}(\Omega) = \frac{1}{2\tau_{imp}}\frac{\Omega\delta(\Omega)}{\pi T}$. Substitution of this into equations (\ref{eq3}) and (\ref{eq4}) gives an additional contribution to the optical scattering rates of $2/\tau_{imp}$ and to the quasiparticle scattering rate of $1/\tau_{imp}$. Thus we recover the well known result that for elastic scattering, the optical rate is twice the quasiparticle rate. We point out that this residual rate $1/\tau^{op}_{imp}(T,\omega)$ remains unrenormalized by the electron-phonon interaction. This is in sharp contrast to the known results about the Drude contribution to $Re[\sigma(T,\omega)]$ which has its optical spectral rate reduced by a factor of $1+\lambda$ where $\lambda$ is the quasiparticle mass renormalization factor. The remaining weight is transferred to the phonon assisted Holstein side bands. In addition the width of the Drude at zero temperature is renormalized to $1/\tau^{op}_{imp}\times 1/(1+\lambda)$.

The optical scattering rate for a correlated system is defined in analogy to the quasiparticle self energy $\Sigma^{qp}(T,\omega)$ through a generalized Drude formula with temperature and energy dependent optical self energy $\Sigma^{op}(T,\omega)$. The infrared conductivity $\sigma(T,\omega)$ is given by
\begin{equation}\label{eq5}
\sigma(T,\omega)=\frac{i}{4\pi}\frac{\Omega_p^2}{\omega-2\Sigma^{op}(T,\omega)}
\end{equation}
where $\Omega_p$ is the plasma energy. We define $1/\tau^{op}(T,\omega)\equiv$ -2$Im[\Sigma^{op}(T,\omega)]$ and the optical renormalized effective mass ($m^{*}_{op}(T,\omega)/m-1)\omega = -2Re[\Sigma^{op}(T,\omega)]$. One can write down explicit formulas for $m^{*}_{op}(T,\omega)$ or get it from $1/\tau^{op}(T,\omega)$ by Kramers-Kronig transform. In terms of the conductivity $\sigma(T,\omega)$ the scattering rate
\begin{equation}\label{eq6}
\frac{1}{\tau^{op}(T,\omega)}=\frac{\Omega_p^2}{4\pi}Re\Big{[} \frac{1}{\sigma(T,\omega)} \Big{]} = \frac{\Omega_p^2}{4\pi} \frac{\sigma_1(T,\omega)}{\sigma_1^2(T,\omega)+\sigma_2^2(T,\omega)}
\end{equation}
where $\sigma(T,\omega) \equiv \sigma_1(T,\omega)+i\sigma_2(T,\omega)$ which allows one to construct the optical scattering rate from a knowledge of the real and imaginary part of the conductivity. In fact only the real part is needed since real and imaginary parts are related by Kramers-Kronig as was the case for the optical self energy itself.

To obtain the electro-phonon spectral density $\alpha^2F(\Omega)$ from the experimental data on $1/\tau^{op}(T,\omega)$ due care must be used to first subtract out the residual scattering contribution. Then a maximum entropy technique can be used to solve the deconvolution problem implied in Eqn. (\ref{eq1}). The method has been widely used particularly in the context of the high T$_c$ cuprates\cite{hwang:2006,carbotte:2011,schachinger:2006,hwang:2006,yang:2009a}.

\begin{figure}[t]
  \vspace*{-1.0 cm}%
  \centerline{\includegraphics[width=4.0 in]{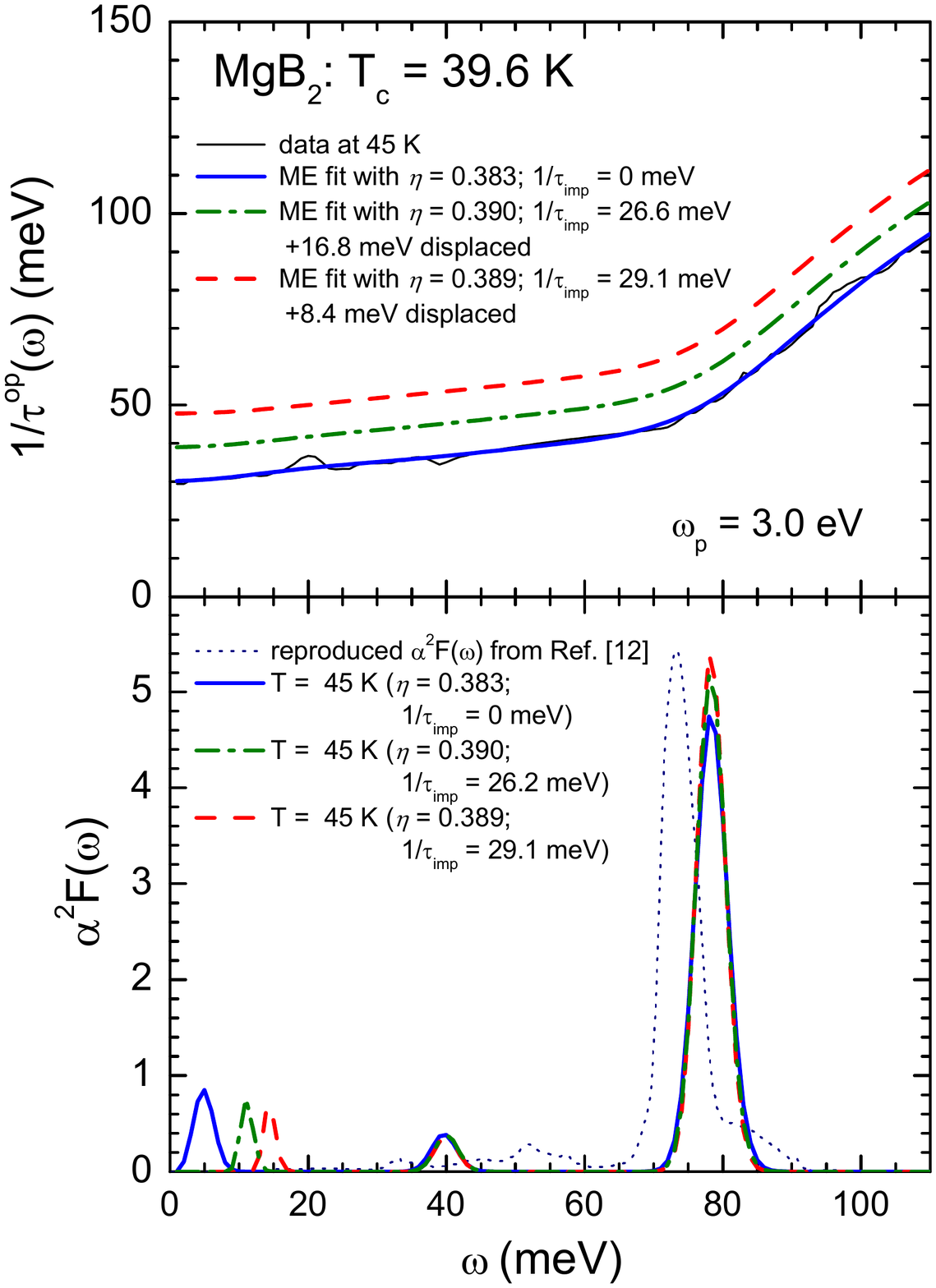}}%
  \vspace*{-0.9 cm}%
\caption{(Color online) The electron-phonon spectral density $\alpha^2F(\Omega)$ (lower frame) recovered from maximum entropy inversion of the optical scattering rate data (upper frame) for MgB$_2$ based on infrared data of reference \cite{tu:2001} at 45 K. The solid blue curve was obtained without first subtracting a residual optical scattering $1/\tau_{imp}$ = 0, subtracting first (dash-dotted green) $1/\tau_{imp}$ = 26.6 meV and for (dashed red) curve subtracting $1/\tau_{imp}$ = 29.1 meV. Second and third curves in top frame displaced upward for viewing ease. The light black curve in the top frame is the data. The black dotted curve in the lower frame is the calculated $\alpha^2F(\Omega)$ from band structure shown in the top frame of Fig. 1 in reference \cite{dolgov:2003}. While the value of $\eta$ is related to the quality of the fit it depends on many factors including the number of data points, their range in energy, and the noise. Also, the value of $\eta$ cannot be made so small that the recovered $\alpha^2F(\Omega)$ itself becomes noisy.}
 \label{fig1}
\end{figure}

\section{Maximum Entropy Inversion}

In Eqn. (\ref{eq1}) what is known is the optical scattering rate $1/\tau^{op}(T,\omega)$ obtained in our case from normal state data at temperature $T =$ 45 K slightly above the superconducting critical temperature $T_c \cong$ 40 K. For a general kernel, $K(T,\omega,\Omega)$ the deconvolution of equation (\ref{eq1}) to recover the spectral density $\alpha^2F(\Omega)$ is ill-conditioned. Here we use a maximum entropy technique\cite{schachinger:2006}. The equation can be discretized $D_{in}(i)=\sum_j K(T,i,j) \alpha^2F(j) \delta\Omega$ where $\delta\Omega$ is the differential increment on the integration over $\Omega_j = j \delta\Omega$ with $j$ an integer. We define a $
\chi^2$ by
\begin{equation}\label{eq7}
\chi^2 = \sum_{i =1}^{N} \frac{[D_{in}(i)-\Sigma(i)]^2}{\epsilon^2_i}
\end{equation}
where $D_{in}(i)$ is the input data, and $\Sigma(i) \equiv \sum_j K(T,i,j)\alpha^2 F(j)\delta\Omega$ is calculated from the known kernel and a given choice of $\alpha^2F(j)$, and $\epsilon_i$ is the error assigned to the data $D_{in}(i)$. Constraints such as positive definiteness for the boson exchange function are noted and the entropy functional
\begin{equation}\label{eq8}
L=\frac{\chi^2}{2}-\eta S
\end{equation}
is minimized with the Shannon-Jones entropy\cite{schachinger:2006}, $S$ given by
\begin{equation}\label{eq9}
S =\int^{\infty}_{0}\Big{[} \alpha^2F(\Omega)-m(\Omega)-\alpha^2F(\Omega)\ln\Big{|} \frac{\alpha^2F(\Omega)}{m(\Omega)}\Big{|}\Big{]} d\Omega.
\end{equation}
The parameter $\eta$ in Eq. (\ref{eq8}) controls how close a fit to the data is obtained. In all cases presented here we have iterated until the average $<\chi^2> = N$ is achieved to acceptable accuracy. The parameter $m(\Omega)$ is here taken to be some constant value on the assumption that there is no a priori knowledge of the functional form of the electron-boson spectral density $\alpha^2F(\Omega)$. Other inversion methods such as singular value decomposition could be used, but here we prefer maximum entropy. Detail comparison of these two methods has appeared in reference \cite{schachinger:2006} where it was found that for Pb both methods give very much the same results. A least square fit could also be employed but it requires making a specific choice for the form of $\alpha^2F(\Omega)$ dependent on a specific number of parameters which are then varied. Here we prefer not to be constraint to a particular functional form.

\section{Numerical Results}

\begin{figure}[t]
  \vspace*{-1.0 cm}%
  \centerline{\includegraphics[width=4.0 in]{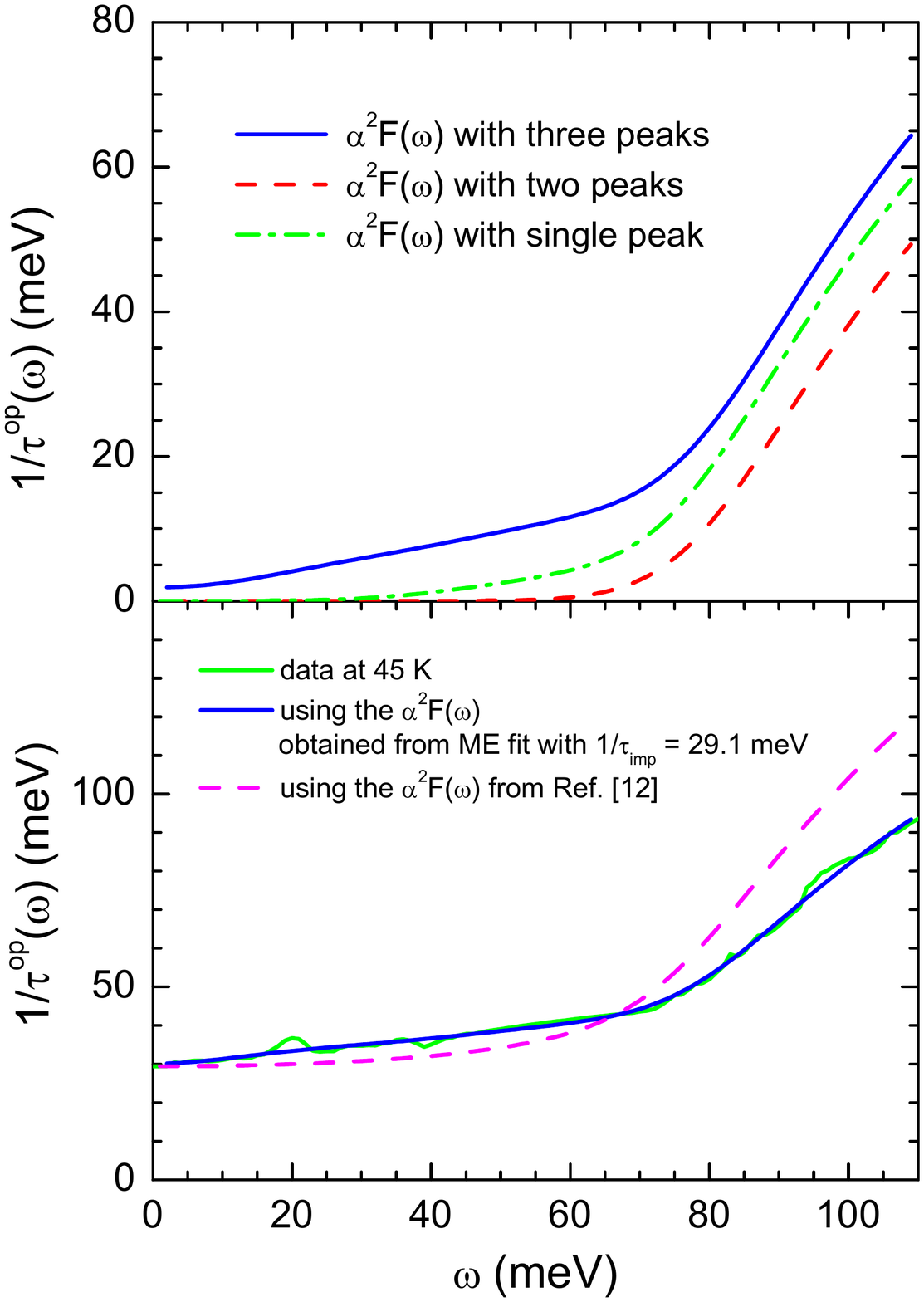}}%
  \vspace*{-0.9 cm}%
\caption{(Color online) (Upper frame) The calculated optical scattering rate based on the recovered $\alpha^2F(\Omega)$ including all three peaks (solid blue), top and middle peak (dash-dotted green) and only top peak (dashed red). (Lower frame) Comparison of optical scattering rate obtained from our $\alpha^2F(\Omega)$ with $1/\tau_{imp} =$ 29.1 meV (solid blue curve) with the result obtained from the $\alpha^2F(\Omega)$ calculated in reference \cite{dolgov:2003} (dashed red curve). The solid light green is the data.}
 \label{fig2}
\end{figure}

\begin{figure}[t]
  \vspace*{-1.0 cm}%
  \centerline{\includegraphics[width=4.0 in]{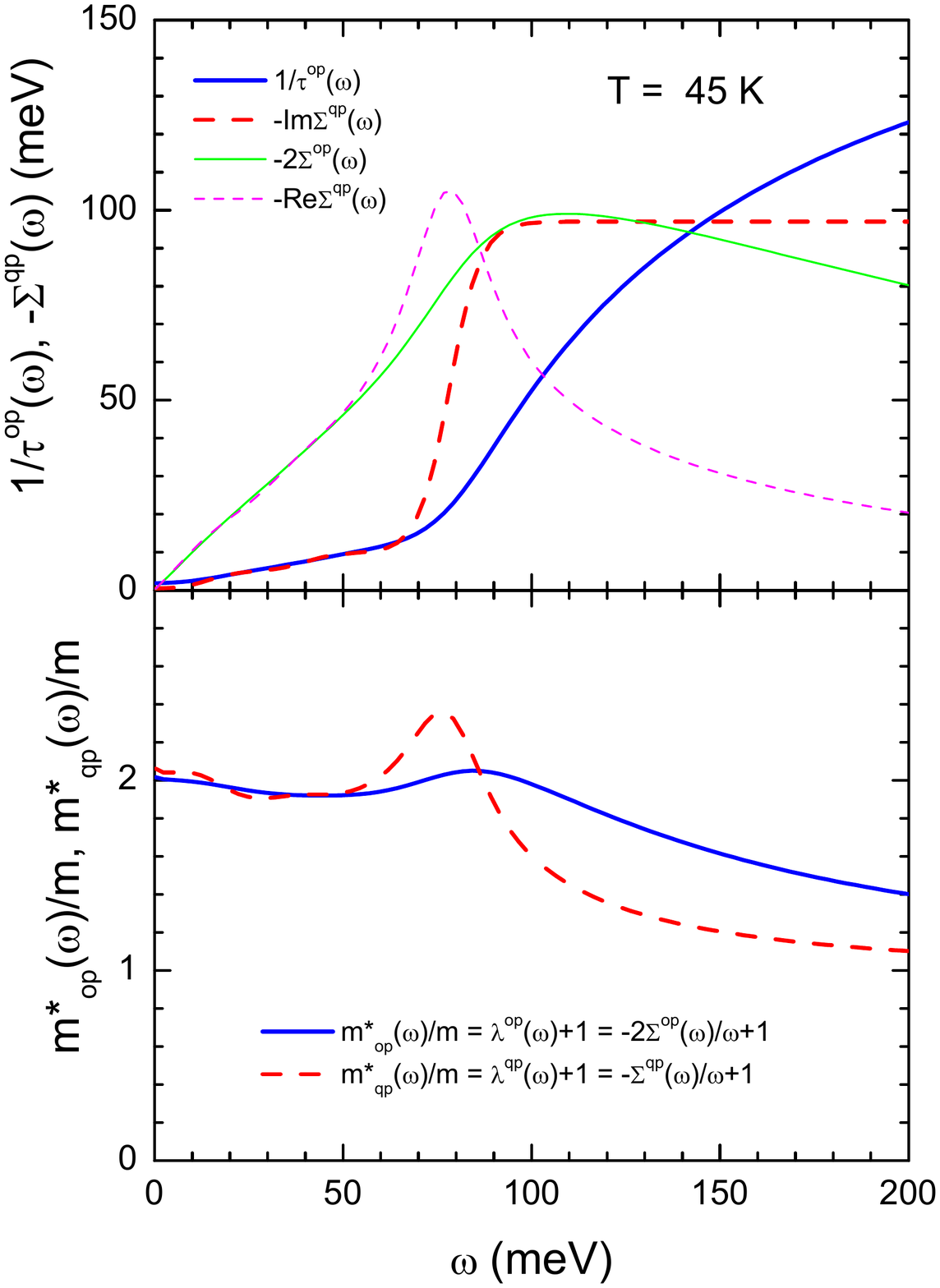}}%
  \vspace*{-0.9 cm}%
\caption{(Color online) The calculated optical self energy (top frame) as a function of $\omega$ compared with the corresponding quasiparticle self energy. Solid blue $1/\tau^{op}(\omega)$, dashed red -$Im\Sigma^{qp}(\omega)$, light solid green -2$\Sigma^{op}_{1}(\omega)$ and light dashed magenta -$Re\Sigma^{qp}(\omega)$. In the lower frame we show the optical effective mass $m^*_{op}(\omega)/m$ (solid blue) which we compare with the quasiparticle effective mass $m^*_{qp}(\omega)/m$ (dashed red).}
 \label{fig3}
\end{figure}

The light black curve in the top frame of Fig. \ref{fig1} gives the optical scattering rate which we got by processing the infrared data of Tu {\it et al.}\cite{tu:2001} on the real and imaginary part of the optical conductivity $\sigma(T,\omega)$ in the normal states of MgB$_2$ at temperature $T =$ 45 K just above the superconducting critical temperature $T_c =$ 39.6 K. According to Eq. (\ref{eq6}) only the real part is needed since $\sigma_2(T,\omega)$ follows by Kramers-Kronig transform. Applying a maximum entropy inversion to obtain from this the spectral density $\alpha^2F(\omega)$ based on the data up to 110 meV, we get the solid blue spectrum shown in the lower frame. The fit to the data that we achieve is very good as we can see from the solid blue curve in the top frame. We emphasize that we get three peaks in our $\alpha^2F(\omega)$ one at low energies centered $\sim$ 5 meV a second centered around 40 meV and a very large peak in comparison to the first two which is centered at 78.3 meV. To obtain this spectrum we used a value of plasma energy $\Omega_p$ of 3.0 eV suggested in the work of Kuzmenko {\it et al.}\cite{kuzmenko:2002}. As can be seen in Eqn. (\ref{eq6}) the value of the optical scattering is directly proportional to the square of $\Omega_p$. Since $1/\tau^{op}(T,\omega)$ in Eq. (\ref{eq1}) is directly proportional to $\alpha^2F(\Omega)$ we see that its overall magnitude can be changed by adjusting the value of the plasma energy.

\begin{table}[t]
  \centering
\begin{tabular}{|c|r|r|r|r|}
  \hline
  $1/\tau_{imp}$ & $\lambda_{peak1}$ & $\lambda_{peak2}$ & $\lambda_{peak3}$ & $\lambda_{total}$ \\ \hline
  \multirow{2}{*}{0 meV}
  & 1.39 & 0.09 & 0.71 & 2.19 \\ \cline{2-5}
   & 63 \% & 4 \% & 33 \% & 100 \% \\ \hline
  \multirow{2}{*}{26.2 meV} & 0.31 & 0.08 & 0.72 & 1.11 \\ \cline{2-5}
   & 28 \% & 7 \% & 65 \% & 100 \% \\ \hline
  \multirow{2}{*}{29.2 meV} & 0.22 & 0.08 & 0.71 & 1.01 \\ \cline{2-5}
   & 22 \% & 8 \% & 70 \% & 100 \% \\
  \hline
\end{tabular}
  \caption{The optical mass enhancement factors. We note that the calculated band spectral enhancement factor $\lambda_{bandST} =$ 1.23.}
  \label{tab1}
\end{table}

In obtaining the solid blue curve we have not included any residual scattering yet we get a good fit to the data. But this requires the relatively large peak at $\sim$ 5 meV which effectively simulates the static residual contribution. If instead we first subtract from the data a constant scattering rate of 26.6 meV we get the dash-dotted dark green curve. Note that the peak at 40 meV and at 78.3 meV have not shifted significantly. It is only the low energy peak which has shifted to $\sim$ 11 meV. If we had used instead 29.1 meV which corresponds to making the contribution from the inelastic scattering negligible at $\omega =$ 0 we would have obtained the dashed red curve in which the lower peak is now shifted to $\sim$ 14 meV. Table \ref{tab1} provides details about the three spectra shown in the lower frame of Fig. \ref{fig1}. The first column is the value of the impurity optical scattering rate $1/\tau^{op}_{imp}$ which was first subtracted. The second column gives the mass enhancement associated with the lowest energy peak $\lambda \equiv 2\int \frac{\alpha^2F(\Omega)}{\Omega}d\Omega$, the third and fourth are contributions from the second peak and third peak respectively and the last column is the total. We note that the $\lambda$ associated with the third peak remains unchanged and that associated with the second peak is also quite stable. This is not true about the lowest energy peak. If no impurity scattering is included in the maximum entropy inversion the large value of $1/\tau^{op}(T,\omega)$ at $\omega =$ 0 for $T =$ 45 K is simulated by this peak. If instead we assign this entire value to residual scattering, then the peak has moved to $\sim$ 14 meV and its $\lambda$ is about 0.22. We conclude from these remarks that the peak at 14 meV is real as is that at 40 meV while the major contribution to the mass enhancement comes from the peak at 78.3 meV and that the total $\lambda$ is $\sim$ 1.01. The last curve (light dotted black) in the bottom frame of Fig. \ref{fig1} given the $\alpha^2F(\Omega)$,  the band structure calculation value of the spectral density, which we read off the heavy solid black curve in Fig. \ref{fig1} of reference \cite{dolgov:2003}. The authors give a value of $\lambda =$ 1.23. Our inversion of optical data agree well with these calculations. The main discrepancy is in the position of the main peak which is shifted to $\sim$ 73 meV in the calculated curve and a smaller lambda. The small peaks that we have found at 14 meV and 40 meV do not aline with the small background seen in the dotted black curve but we believe that they are real.

 The two small peaks located at 14 meV and 40 meV can be related directly to measured features in the scattering rate data as we illustrate in Fig. \ref{fig2}. The upper frame shows the optical scattering rate that is obtained from our model spectral density $\alpha^2F(\Omega)$. The solid blue curve includes all three peaks, the dash-dotted green leaves out the lower peak at $\sim$ 14 meV and the red dashed leaves out both the $\sim$ 14 meV and the 40 meV peaks. It is clear that both these peaks are needed to fit the data which basically fall on the solid blue curve. On the other hand the dotted black curve in the lower frame of Fig. \ref{fig1} which gives the calculated $\alpha^2F(\Omega)$ of band theory does not fit the data as can be seen in the lower frame. The optical scattering rate obtained in this way is given by the magenta dashed curve. Not only does it not rise enough as compared  with the data (light green curve) in the region below 60 meV but it also overshoots the data above 60 meV. Formula (\ref{eq6}) for the optical scattering rate involves the total conductivity $\sigma(\omega)$ and in the case of multiple bands, as we have in MgB$_2$, a sum needs to be taken before $1/\tau^{op}(T,\omega)$ is constructed. However if in the energy range of interest one band is much more resistive than the other one would get from our maximum entropy inversion the spectral density $\alpha^2F(\Omega)$ associated with the dominant band rather than a composite Kuzmenko {\it et al.}\cite{kuzmenko:2002} have provided evidence that the three dimensional band may be close to weak localization and consequently we expect that we are involved mainly with the two dimensional $\sigma$-band. Consequently we have compared in the lower frame of Fig. \ref{fig1} our results (dashed red) for $\alpha^2F(\Omega)$ with the computed spectrum for the two dimensional $\sigma$-band (dotted red). We note that the effective width of the Drude with a $\lambda$ of 1.01 is the bare residual scattering rate here $\sim$ 30 meV divided by 1+$\lambda$ which is $\sim$ 15 meV in reasonable agreement with the data of Tu {\it et al.}\cite{tu:2001} which gives $\sim$ 20 meV somewhat higher. All the discrepancy noted could be due to some uncertainty in our knowledge of the plasma frequency for the $\sigma$-band or perhaps there is a small contribution to $1/\tau^{op}(T,\omega)$ from the $\pi$-band.

\begin{figure}[t]
  \vspace*{-0.5 cm}%
  \centerline{\includegraphics[width=4.5 in]{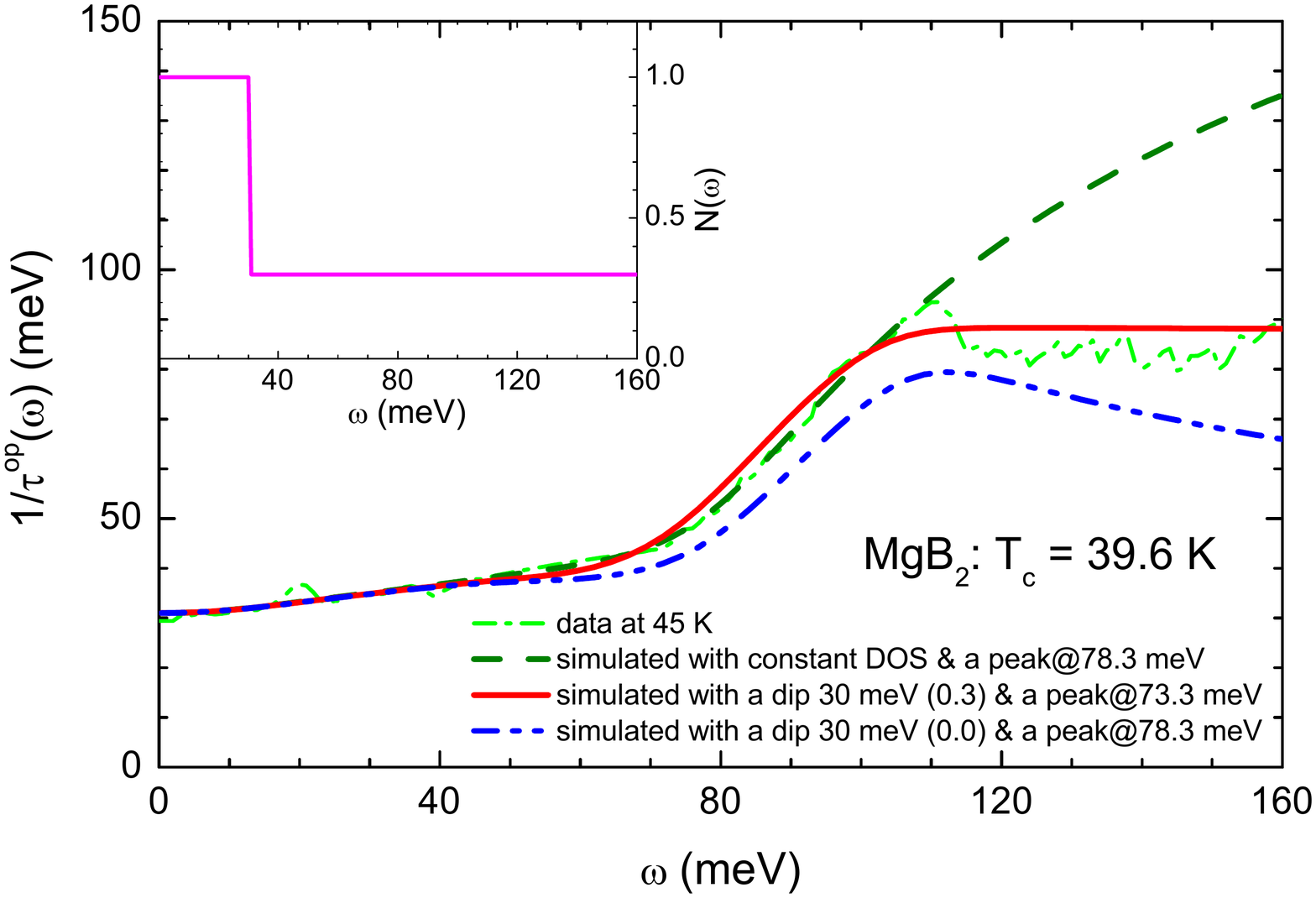}}%
  \vspace*{-0.9 cm}
\caption{(Color online)  We show additional results obtained in various way with an extended energy range to 160 meV. The data is the green dash-dotted curve, the dashed dark green the calculated optical scattering rate using the $\alpha^2F(\Omega)$ of Fig. \ref{fig1} top frame (including all three peaks). The other two curves include an energy dependent density of states as shown in the inset (magenta curve) with a drop at 30 meV. For the solid red curve the drop is to 0.3 while for the dash-double dotted blue curve it is to 0.0. In the solid red curve the high energy peak in $\alpha^2F(\Omega)$ has been shifted down from 78.3 meV to 73.3 meV.}
 \label{fig4}
\end{figure}

It is interesting to note the difference in behavior between the optical self energy and the better known quasiparticle self energy. The first is characteristic of the optical conductivity which deals with a two-particle Green's function while the second is a one-particle property so that a priori there is no compelling reason why they should be the same. In Fig. \ref{fig3} we compare in the upper frame the optical scattering rate (solid blue curve) with its quasiparticle counterpart dashed red curve. We see that $1/\tau^{qp}(T,\omega)$ rises sharply around $\omega =$ 70 meV and rapidly becomes constant while the rise in blue curve is much more gradual and in fact is still rising at 200 meV well above the maximum phonon energy in $\alpha^2F(\Omega)$ which is $\leq 85$ meV. The peak in the real part of the quasiparticle self energy (light dashed magenta curve) is also much sharper than is the peak in the real part of the optical self energy (light solid green curve) as we could have expected since these are related to their imaginary part by Kramers-Kronig transform. Note however that below 50 meV quasiparticle and optical quantities are quite close to each other. In the lower frame of Fig. \ref{fig3} we show additional results for the optical (solid blue) and quasiparticle (dashed red) effective masses. They agree well at $\omega \rightarrow 0$, but differ beyond this region. At $T =$ 0 $\lambda^{op}$ and $\lambda^{qp}$ would be the same and would also be equal to the spectral lambda, $\lambda$.

So far we have considered in our maximum entropy fits to the data only the energy range up to 110 meV. We have also assume that the electronic density of states can be taken as a constant in this range. Around 120 meV and beyond however the data for $1/\tau^{op}(T,\omega)$ takes on a very different character. Rather than keep rising it has a small peak and then becomes rather flat. This is shown in Fig. \ref{fig4} as the dash-dotted green curve. The dashed dark green curve is theory using the $\alpha^2F(\Omega)$ that we obtained from our analysis of data below 110 meV. It is in striking disagreement with the data. A possible explanation is that there is a drop in the electronic density of states in this region and that we should be using Eq. (\ref{eq2}) with a non-constant $\tilde{N}(\omega)$ rather than the simpler Eq. (\ref{eq3}). We show in the inset a model for the energy dependence of $\tilde{N}(\omega)$. With this model we obtain for $1/\tau^{op}(T,\omega)$ the blue dash-dot-dotted curve which shows a peak around 110 meV and then a drop at higher energies. Here we note that the onset energy of the drop in the optical scattering rate is the peak energy (78.3 meV) in $\alpha^2F(\Omega)$ plus the sharp drop energy (30 meV) in $\tilde{N}(\omega)$\cite{hwang:2013a}. To obtain this curve we have used to $\alpha^2F(\Omega)$ obtained with $1/\tau_{imp} =$ 29.1 meV in Fig. \ref{fig1} bottom frame. If however we shift the energy of the high $\omega$ peak in this distribution function from its position at 78.3 meV to 73.3 meV which is the energy of the peak predicted in band structure calculation we get the solid red curve which provides a reasonable understanding of the data. We emphasize that we are not attempting a tight fit to the data in this figure. Rather we simply want to illustrate that including some energy dependence in the density of states $\tilde{N}(\omega)$ allows us to get a qualitative understanding of the data in the range 100 to 160 meV. Another possibility is that in this phonon energy range other bands start to become important including interband transitions. There is however no sign that these play a role below 110 meV which is the range used to obtain our spectral density $\alpha^2F(\Omega)$.

\section{Conclusions}

We have extracted from normal state optical data at a temperature of 45 K the electron-phonon spectral density $\alpha^2F(\Omega)$ of MgB$_2$. We find good agreement with the calculated spectral density obtained from first principle density functional LDA band structure computations. We identify the recovered spectral density as that for the two dimensional sigma band with no evidence for a significant contribution from a second band. This observation supports the idea that the three dimensional $\pi$ band may be highly resistive possibly close to the weak localization regime as suggested in the work of Kuzmenko {\it et al.}\cite{kuzmenko:2002}. This would not be inconsistent with the observation of a small superconducting gap associated with this band. Because the gap is s-wave and isotropic, Anderson's theorem guarantees that intraband elastic impurity scattering drops out entirely from the gap equation for that band.

In our maximum entropy reconstructions of the optical scattering rate data we limited the phono energy range to 110 meV, a value chosen to fall above the maximum phonon energy in MgB$_2$. Above this energy in the range 110 to 160 meV the experimental scattering rate shows unexpected leveling off which cannot be understood in a conventional one band phonon model. In such a model the optical scattering rate would continue to increase even when there is no further boson structure and $\alpha^2F(\Omega)$ has become zero. This is in contrast to the quasiparticle scattering rate which saturates above the highest boson energy. A possible explanation for the flattening out is the existence of a drop in the electronic density of states which reduces the scattering in this energy range below the value it would have if the DOS was constant. It could also be evidence that additional bands make a nonnegligible contribution which modifies the scattering rate in this energy range.

\ack

JH acknowledges financial support from the National Research Foundation of Korea (NRFK Grant No. 20100008552). This work also was supported by Mid-career Researcher Program through NRF grant funded by the Ministry of Education, Science \& Technology (MEST) (No. 2010-0029136). JPC was supported by the Natural Science and Engineering Research Council of Canada (NSERC) and the Canadian Institute for Advanced Research (CIFAR).

\section*{References}
\bibliographystyle{unsrt}
\bibliography{bib}

\end{document}